\documentclass[10pt,twoside]{article}
\usepackage{hyperref}
\usepackage{amssymb}
\usepackage[mathscr]{eucal}
\usepackage{eufrak}
\usepackage{amsmath}
\usepackage{mathrsfs}

\def\intl#1{\int\limits_{#1}}
\def\intll#1#2{\int\limits_{#1}^{#2}}
\def\dm{|\hskip-0.05cm|}

\def\displ{\displaystyle}
\def\VSE{\vspace{6pt}\\&\displ }
\def\VS{\vspace{6pt}\\\displ }
\def\rf#1{{\rm(\ref{#1})}}
\def\chiu{\hfill$\displaystyle\vspace{4pt}
\underset\Box\null$\par}
\def\Pr{{\bf Proof. }}

\def\R{\Bbb R}

\def\à{\`{a}}
\def\è{\`{e}}
\def\ì{\`{i}}
\def\ù{\`{u}}
\def\ò{\`{o}}
\def\é{\'{e}}

\def\vep{\varepsilon}

\def\be{\begin{equation}}
\def\ba{\begin{array}}
\def\ea{\end{array}}
\def\ee{\end{equation}}
\def\vs1{\vspace{1ex}}

\def\ov{\overline}

\font\sc=cmcsc10
\title{A remark on the partial regularity of a suitable weak solution to the Navier-Stokes Cauchy problem}
\author{\sc F. Crispo and P. Maremonti
\thanks{
Dipartimento di Matematica e Fisica, Seconda
Universit\`{a} degli Studi di
 Napoli, via Vivaldi 43, 81100 Caserta,
 Italy.
francesca.crispo@unina2.it;
paolo.maremonti@unina2.it}}
\date{}
\begin{document}
\maketitle \noindent{\bf Abstract} - {\small   Starting from the partial regularity results for suitable weak solutions to the Navier-Stokes Cauchy problem by Caffarelli, Kohn and Nirenberg \cite{CKN},   as a corollary,  under suitable assumptions of local character  on the initial data, we prove a  behavior in time of the $L^\infty_{loc}$-norm of the solution in a neighborhood of $t=0$. The behavior is the same as for the resolvent operator associated to the Stokes operator. 
Besides its own interest, the result is a main tool to study the spatial decay estimates of a suitable weak solution, performed in  paper \cite{CMSD}. }
 \par\noindent
 \vskip -0.7true cm\noindent
\newcommand{\red}{\protect\bf}
\renewcommand\refname{\centerline
{\red {\normalsize \bf References}}}
\newtheorem{ass}
{\bf Assumption}[section]
\newtheorem{defi}
{\bf Definition}[section]
\newtheorem{tho}
{\bf Theorem}[section]
\newtheorem{rem}
{\sc Remark}[section]
\newtheorem{lemma}
{\bf Lemma}[section]
\newtheorem{coro}
{\bf Corollary}[section]
\newtheorem{prop}
{\bf Proposition}[section]
\renewcommand{\theequation}{\thesection.\arabic{equation}}
\setcounter{section}{0}
\numberwithin{equation}{section}
\section{Introduction}\label{intro}
In this note we study the  partial regularity
of suitable weak solutions to the
Navier-Stokes initial value
problem:
\be\label{NS}\ba{l}v_t+v\cdot
\nabla v+\nabla\pi_v=\Delta
v,\;\nabla\cdot
v=0,\mbox{ in }(0,T)\times\R^3,\\
v(0,x)=v_\circ(x)\mbox{ on
}\{0\}\times\R^3.\ea\ee In system
\rf{NS} $v$ is the kinetic field,
$\pi_v$ is the pressure field,
 $v_t:=
\frac\partial{\partial t}v$  and
 $v\cdot\nabla v:=
v_k\frac\partial{\partial x_k}v$.
In order to highlight the main ideas we assume zero body
force, as well as we restrict ourselves to the Cauchy problem. However, the initial boundary value problem will be considered in a forthcoming paper.

The symbol
$\mathscr C_0(\R^3 )$ stands for  
the subset of $C_0^\infty(\R^3 )$
whose elements are divergence
free. We set
$J^2(\R^3)\!:=$completion of
$\mathscr C_0(\R^3)$ with respect the $L^2$-norm,   
and $J^{1,2}(\R^3)\!:=$ completion of 
$\mathscr C_0(\R^3)$  with respect to  the  $W^{1,2}(\R^3)$-norm.  

 We set
$(u,g):=\intl{\!\!\R^3
} u\cdot gdx\,.$

In the present note, assuming $v_\circ\in J^2(\R^3 )$, 
and, for $x_0\in \R^3$ and $R_0>0$,   \be\label{IC}\mathscr E_0(x_0,R_0):=\mbox{ess}\hskip-0.1cm\sup_{\hskip-0.4cmB(x_0,R_0)}\dm
v_\circ\dm_{w(x)}:=\dm\big(\intl{\R^3}\frac{|v_\circ(y)|^2}{|x-y|}
dy\big)^\frac12\dm_{L^\infty(B(x_0,R_0))}
\ee 
small in a suitable sense, where the smallness is independent
of $B(x_0,R_0)$ and of $v_\circ$, we prove, in a neighborhood of $t=0$, a time-weighted estimate of the $L^\infty_{\ell oc}$-norm of a suitable weak solution (hence local regularity for all $t>0$). 

 Our result has to be put in the context of the general results obtained by Caffarelli-Kohn-Nirenberg in \cite{CKN}, that are crucial for our aims, although other results related to sufficient conditions for regularity could be employed (see e.g. \cite{Farwig}, \cite{LS}, \cite{L}, \cite{S}, \cite{V}).  
 
  To
better state our main result, we
introduce some definitions and notation. This is done following as much as possible the ones in \cite{CKN}.\begin{defi}\label{WS}{\sl  A pair $(v,\pi_v)$, such that $v:(0,T)\times\R^3 \to\R^3$ and $\pi_v:(0,T)\times\R^3 \to \R$, is said a weak solution to problem {\rm\rf{NS}} if
\begin{itemize}\item [i)] for all $T>0$,
 $v\in  L^2(0,T; J^{1,2}
(\R^3 ))$ and $\pi_v\in L^\frac53((0,T)\times\R^3)$, $$\dm v(t)\dm_2^2+2\intll st \dm\nabla v(\tau)\dm_2^2d\tau\leq \dm v(s)\dm_2^2,\; \mbox{ for all } t\geq s, \mbox{ for } s=0 \mbox{ and a.e. in } s\geq 0\,,$$\item
[ii)] $\displ\lim_{t\to0}\dm
v(t)-v_\circ\dm_2=0$,\item[iii)]
for all $t,s\in(0,T)$,  the pair $(v,\pi_v)$
satisfies the  equation:
\newline \centerline{$\displ\intll
st\Big[(v,\varphi_\tau)-(\nabla
v,\nabla
\varphi)+(v\cdot\nabla\varphi,v)+(\pi_v,\nabla
\cdot\varphi)\Big]d\tau+(v(s),\varphi
(s))=(v(t),\varphi(t))$,}
\newline\centerline{
 for all $\varphi\in C^1_0([0,T)\times\R^3 )$.}
\end{itemize}}\end{defi} 
\begin{defi}\label{SWS}{\sl A pair $(v,\pi_v)$ is said a suitable weak solution if it is a weak solution in the sense of the Definition\,\ref{WS} and, moreover,
\be\label{SEI}\ba{l}\displ\intl{\R^3}|v(t)|^2\phi(t)dx+2\intll st\intl
{\R^3}|\nabla v(\tau)|^2\phi\, dxd\tau\leq \intl{\R^3}|v(s)|^2\phi(s)dx\VS\hskip 2,5cm+\intll st\intl{\R^3}|v|^2(\phi_\tau+\Delta\phi)dxd\tau+\intll st\intl{\R^3}(|v|^2+2\pi_v)
v\cdot\nabla\phi dxd\tau,\ea\ee for all $t\geq s$, for $s=0$ and a.e. in $s\geq 0$, and for all 
nonnegative $\phi\in C_0^\infty(\R\times\R^3)$.}\end{defi} In \cite{CKN} and \cite{scheffer1} the following existence result is proved:
\begin{tho}\label{EXCKN}{\sl For all $v_\circ\in J^2(\R^3)$ there exists a suitable weak solution.}\end{tho}
Concerning the regularity of a suitable weak solution,  we begin by 
giving the following
\begin{defi}\label{RP}{\sl We say
that $(t,x)$ is a singular point
for a weak solution $(v,\pi_v)$ if
$v\notin L^\infty$ in any
neighborhood of $(t,x)$; the
remaining points, where $v\in
L^\infty(I(t,x))$ for some
neighborhood  $I(t,x)$, are
called regular. }\end{defi}
We introduce the parabolic cylinders 
$$Q_r(t,x):=\{(\tau,y):t-r^2<\tau<t\mbox{ and }|y-x|<r\}, $$ 
and 
$$Q_r^*(t,x):=\{(\tau,y):t-\frac78r^2<\tau<t+\frac18r^2\mbox{ and }|y-x|<r\},$$ that is $Q_r^*(t,x)=Q_r(t+\frac 18r^2,x)$.
Moreover, for $r\in (0,t^\frac12)$, we set
$$M(r):=r^{-2}\int\hskip-0.2cm\intl{Q_r}(|v|^3+|v||\pi_v|)dyd\tau+r^{-\frac{13}4}\intll{t-r^2}t\big(\hskip-0.15cm\intl{|x-y|<r}\hskip-0.2cm|\pi_v|dy\big)^\frac54d\tau\,,$$
where, for simplicity, we suppress the dependence on $(t,x)$. 

 In \cite{CKN}, to achieve the regularity of a suitable weak solution two sufficient conditions are given. The first one is Proposition\,1 (or Corollary\,1) in \cite{CKN}:
\begin{prop}\label{RCKN}{\sl Let $(v,\pi_v)$ be a suitable weak solution in some parabolic cylinder $Q_r(t,x)$. There exist $\vep_1>0$ and $c_0>0$ independent of $(v,\pi_v)$ such that, if $M(r)\leq \vep_1$, then
\be\label{RCKNI}|v(\tau,y)|\leq c_1^\frac12r^{-1},\mbox{ a.e. in }(\tau,y)\in Q_{\frac r2}(t,x),\ee where $c_1:=c_0\vep_1^{\frac23}$. In particular, a suitable weak solution $v$ is regular in $Q_{\frac r2}(t,x)$. }\end{prop}
Since in the statement of Proposition\,\ref{RCKN} it is required $M(r)\leq \vep_1$ for all $r\in(0,t^\frac12)$, and since $M(r)$ is in integral form, we do not prejudice  the problem  giving the definition of $M(r)$ for $r\in(0,t^\frac12]$.
 
  The second sufficient condition is Proposition\,2 in \cite{CKN}: 
\begin{prop}\label{RICKN}{\sl There is a constant $\vep_3>0$ with the following property. If $(v,\pi_v)$ is a suitable weak solution in some parabolic cylinder $Q_r^*(t,x)$ and 
$$\limsup_{r\to0}r^{-1}\int\hskip-0.2cm\intl{Q_r^*}
|\nabla v|^2dyd\tau\leq \vep_3\,,$$ then $(t,x)$ is a regular point.}\end{prop}
In \cite{CKN}, the above propositions are crucial to deduce the regularity for a suitable weak solution.    In their applications, in particular Theorem\,C and Theorem\,D, and Corollary on p.820, the regularity results express on the geometry of regular points, but either a continuous dependence on the initial data and a behavior in a neighborhood of $t=0$ do not seem an immediate consequence. Of course, the same difficulties arise from the results of \cite{Farwig}, \cite{LS}, \cite{L}, \cite{S} and \cite{V}.

The following Theorem\,\ref{CT}, which is the 
chief result of this note, is related to the pointwise continuous dependence of the null solution, in the framework of partial regularity for a suitable weak solution. \par From now on we will assume that $v_\circ\in J^2(\R^3)$ and, in the light of Theorem \ref{EXCKN},  denote by $(v,\pi_v)$ a corresponding suitable weak solution. 

\begin{tho}\label{CT}{\sl Let $(v,\pi_v)$ be a suitable weak solution. Then there exist absolute constants $\vep_1$, $C_1$ and $C_2$ such that if 
\be\label{th}\;
C_1\mathscr E_0( x_0,R_0)<1\ \mbox{ and }\ C_2(\mathscr E_0^3+\mathscr E_0^\frac52)\leq\vep_1,\ee
then 
\be\label{thh}|v(t,x)|\leq
c(\mathscr E_0^3+\mathscr E_0^\frac52)^\frac13t^{-\frac12},\ee provided that $(t,x)$ is a Lebesgue point with $\dm v_\circ\dm_{w(x)}<\infty$ and $x\in B(x_0,R_0)$.}
\end{tho}
In the above statement $\vep_1$ is the same as in Proposition\,\ref{RCKN}.

The proof of the theorem is based on Proposition\,\ref{RCKN}. This is also made in \cite{CKN}, but our approach to the proposition is different. Indeed, we prove a weighted energy relation (the norm is $\dm v(t)\dm_{w(x)}$) which holds for $t>0$, provided that \rf{th}$_1$ holds (see estimate \rf{WERI} in Proposition\,\ref{WER}). 

 Theorem\,\ref{CT} is of primary importance as a premise to another paper, \cite{CMSD}, concerning the space-time decay of suitable weak solutions, provided that the same behavior is assumed on the initial data.\vskip0.2cm We end the introduction with few remarks. 
\vskip0.2cm 
Our  partial regularity result states
that under the smallness assumption \rf{th}
almost all $(t,x)\in (0,T)\times
B(x_0,R_0)$ are points of regularity for a
suitable weak solution $v$. In particular, if
$B(x_0,R_0)\equiv \R^3$, then we get a new sufficient condition
for the existence of a global smooth solution.
\par
 Condition \rf{IC} becomes a norm.   We remark that for all $v_\circ\in
J^2(\R^3)$, by virtue of the Hardy-Littlewood-Sobolev theorem,  we have $\dm
v_\circ\dm_{w(x)}<\infty$ almost everywhere in
$x\in\R^3$.  Hence, any data in $L^2$ inherently
defines a   functional that we can assume as norm. 
Condition \rf{th} is just the smallness of the norm $\dm \dm
v_\circ\dm_{w(x)}\dm_{L^\infty(B(x_0,R_0))}$.
We can find several sufficient conditions on $v_\circ\in J^2(\R^3)$ such that assumption \rf{th} is verified. For example, if we consider the assumption in \cite{CKN} (Corollary on p. 820), that is $v_\circ\in W^{1,2}(\R^3-B_R)$, then $\dm v_\circ\dm_{w(x)}$ is a continuous function of $x$ that we can make small outside a suitable ball of radius $R\hskip0.05cm'\geq R$.
\par
 The partial regularity has
a local character in the sense that 
condition \rf{th} can be not satisfied on
$\R^3-B(x_0,R_0)$, and the regularity is ensured a.e. in
$ B(x_0,R_0)$.
Another feature that expresses the local character of the result is the following: the pointwise
behavior of our solution $v$ is given in a neighborhood of
$(0,x_0)$. 
As far as we know, this property, which  is as the one
of the solutions to the Stokes problem,  is new.   
 \par
 Of course estimate \eqref{thh} also gives a pointwise asymptotic behavior of the solution for large $t$. However such a behavior is not optimal under the assumption $v_\circ\in J^2(\R^3)$. Indeed  for a 
 suitable weak solution the pointwise asymptotic behavior is of the kind $O(\dm v_\circ\dm_2t^{-\frac34})$, according to
  the fact that a weak solution becomes smooth for $t>T_0$, where $T_0$
   is connected with the $L^2$-norm of $v_\circ$, and the behavior is 
   governed by the $L^2$-norm of the initial data (see \cite{MCMP}).
 \par
 We conclude by observing that   a weaker 
result   can be deduced  for the initial
boundary value problem. In this case the local regularity is  far from the boundary. The result will be object of a next paper. We attack the problem by using the arguments developed in \cite{MRP} and  new estimates on the pressure field deduced in \cite{J}.       
\section{Partial regularity results}
Firstly we
recall 
some 
results fundamental for our
aims.  \begin{lemma}\label{WI}{\sl Suppose that
$|x|^\beta u\in L^2(\R^3)$ and
$|x|^\alpha \nabla u\in L^2(\R^3)$.
Also
\begin{itemize}\item[i)] $r\geq2$, $\gamma+\frac3r>0$, $\alpha+\frac32>0$, $\beta+\frac32>0$, and $a\in[\frac12,1]$,
\item[ii)]
$\gamma+\frac3r=a(\alpha+\frac12)+(1-a)(\beta+\frac32)$
(dimensional balance), \item[iii)]
$a(\alpha-1)+(1-a)\beta\leq\gamma\leq
a\alpha+(1-a)\beta$.
\end{itemize}
Then, with a constant $c$
independent of $u$,
 the following inequality holds:
\be\label{WII}\dm |x|^\gamma
u\dm_r\leq c\dm |x|^\alpha\nabla
u\dm_2^a\dm |x|^\beta
u\dm^{1-a}_2.\ee
 }\end{lemma}
\Pr See \cite{CKN}
Lemma\,7.1\,.\chiu
 \begin{lemma}\label{CZW}{\sl Assume that $\mathbb K$ is
a singular bounded transformation from $L^p$ into $L^p$,
$p\in(1,\infty)$,  of
Calder\'on-Zigmund kind. Then,
$\mathbb K$ is also a bounded
transformation from $L^p$ into $L^p$
with respect to the measure
$(\mu+|x|)^\alpha dx,\mu\geq 0,$ provided that
$\alpha\in
(-n,n(p-1))$.}\end{lemma}\Pr See
\cite{ST} Theorem\,1.\chiu
\par For the reader's convenience, here we restate Proposition\,\ref{RCKN}, that is Corollary 2 in \cite{CKN} in a slightly different form, more convenient for our aims:
\begin{tho}\label{CKNT}{\sl Let $(v,\pi_v)$ be a suitable weak solution in some parabolic cylinder $Q_r(t,x)$. There exist $\vep_1>0$ and $c_0>0$ independent of $(v,\pi_v)$ such that, if $M(r)\leq \ov\vep_1\leq\vep_1$, then
\be\label{CKNRI}|v(\tau,y)|\leq c_1^\frac12r^{-1},\mbox{ a.e. in }(\tau,y)\in Q_{\frac r2}(t,x),\ee  where $c_1:=c_0\ov\vep_1^{\frac23}$.}\end{tho}
\Pr For the regularity result see Corollary\,1  on p.776 of \cite{CKN}, and for formula $c_1=c_0\ov\vep_1^\frac23$ see (4.4) on p.789  of \cite{CKN}.\chiu
Theorem\,\ref{CT} is proved employing Theorem\,\ref{CKNT}. So that our task is reduced to prove that, under the assumption \rf{th}, for all suitable weak solutions $(v,\pi_v)$ the following holds  $$M(r)\leq c(\mathscr E_0^3+\mathscr E_0^\frac52)\,.$$
\par To this end we will prove
\begin{prop}\label{WER}{\sl Let be $C_1\mathscr E_0<1$. Then there exists a set $D$ and a suitable weak solution $(v,\pi_v)$ such that $meas(B(x_0,R_0)-D)=0$ and
\be\label{WERI}\ba{l}\displ\dm v(t)\dm_{w(x)}^2 +c(\mathscr E_0)
\intll0t\dm\nabla u(\tau)\dm^2_{w(x)}d\tau<c\dm v_\circ\dm_{w(x)}^2\,,\\\hskip 4cm\mbox{ for all }t>0\mbox{ and }x\in D,\ea\ee
where $c(\mathscr E_0)>1$. }\end{prop}
We postpone the proof of the proposition to the next section and now we deduce the above implication.\begin{lemma}\label{RFP}{\sl Assume that $(v,\pi_v)$ is a suitable weak solution. Then the pressure field admits the following representation formula
\be\label{PI}\pi_v(t,x)=-D_{x_i}D_{x_j}\intl{\R^3}\mathcal E(x-y)v^i(y)v^j(y)dy\,,\mbox{ a.e. in }(t,x)\in (0,\infty)\times\R^3.\ee}\end{lemma}
\Pr  Since $\pi_v\in L^\frac53((0,T)\times\R^3)$ and $v\in L^2(0,T;J^{1,2}(\R^3))$, there exists a zero Lebesgue measure set $N$ such that, for all $t\in (0,T)-N$, $\dm\pi_v(t)\dm_\frac53+\dm v(t)\dm_{1,2}<\infty$. 
 To prove (\ref{PI}) we start from the formulation of  weak solution to the Navier-Stokes Cauchy problem evaluated for $t\in (0,T)-N$ and for all $\varphi\in C_0^\infty([0,T)\times\R^3)$:
$$(v(t+\delta),\varphi(t+\delta))-(v(t),\varphi(t))=\intll t{t+\delta}\big[(v,\varphi_\tau+\Delta \varphi)+(v\cdot\nabla\varphi,v)+(\pi_v,\nabla \cdot \varphi)]d\tau.$$
In particular setting $\varphi(\tau,x):=h(\tau)\nabla \psi(x)$, where $
h(\tau):=\left\{\ba{l}1\mbox{ for }\tau\leq t+\delta\,,\\0\mbox{ for }\tau\geq t+2\delta\,,\ea\right.$ is a smooth cut-off function and $\psi(x)\in C_0^\infty(\R^3)$, since $v$ is divergence free, we get
$$\intll t{t+\delta}\big[(v\cdot\nabla\nabla \psi,v)+(\pi_v,\Delta\psi)\big]
 d\tau=0.$$
 Multiplying by $\frac1\delta$, in the limit as $\delta\to 0$, we deduce
 \be\label{PII}(\pi_v,\Delta \psi)=-(v\otimes v,\nabla\nabla \psi),\mbox{ a.e. in }t>0,\mbox{ for all }\psi\in C_0^\infty(\R^3)\,.\ee
A solution of equation (\ref{PII}) is given by 
$$\ov\pi=-D_{x_i}D_{x_j}\intl{\R^3}\mathcal E(x-y)v^i(t,y)v^j(t,y)dy\,.$$ Since, for $t\in (0,T)-N$, we have $v\in J^{1,2}(\R^3)\subset L^\frac{10}3(\R^3)$,
by virtue of Calder\'on-Zigmund theorem we get $\ov\pi\in L^\frac53(\R^3)$.
Hence, we deduce, almost everywhere in $t>0$,
$$(\pi_v-\ov\pi,\Delta \psi)=0,\mbox{ for all }\psi\in C_0^\infty(\R^3),$$
with $\pi_v-\ov\pi\in L^\frac53(\R^3)$, so that $\pi_v\equiv\ov\pi$ and \rf{PI} is achieved.  \chiu
\par If \rf{WERI} holds, then, with the aid of the previous lemma,  we are able to prove the following 
\begin{lemma}\label{RLI}{\sl Assume that (\ref{WERI}) holds for a suitable weak solution $(v,\pi_v)$. Then
$M(r)\leq C_2(\mathscr E_0^3+\mathscr E_0^\frac52)$, for $r^2\in(0,t)$.}\end{lemma}\Pr  By virtue of our assumption \rf{WERI}, by virtue of representation formula \rf{PI} and Lemma \ref{CZW},  a.e. in $t>0$, we   get that 
\be\label{PIII} \dm \pi_v(t) |x-y|^{-\frac43}\dm_{\frac32}\leq c
\dm |v(t)| |x-y|^{-\frac23}\dm_{3}^2\,.\ee
Applying H\"older's inequality, from \rf{PIII} and from Lemma\,\ref{WI} we get
\be\label{PIV}\ba{ll}\displ r^{-2}\!\!\!\intll{t-r^2}t\,\intl{|x-y|<r}\!\!\!\!\Big[|v|^3+|v||\pi_v|\Big]dyd\tau\hskip-0.3cm&\displ\leq\! c\!\intll{t-r^2}t\!\!\Big[\!
\dm\frac{ v(\tau)}{ |x-y|}\null_{_{\!\!\frac23}}\!\dm_{3}^3+
\dm \frac{v(\tau)}{ |x-y|}\null_{_{\!\!\frac23}}\!\dm_{3}
\dm \frac{\pi_v(\tau) }{|x-y|}\null_{_{\!\!\frac43}}\!\dm_{\frac32}\!\Big]d\tau\VSE\leq \!c\!\intll{t-r^2}t\!\dm \frac{v(\tau)}{|x-y|
}\null_{\null_{\frac12}}\!\dm_2 \dm\frac{\nabla v(\tau)}{|x-y|}\null_{\null_{\!\frac12}}\!\dm_2^2d\tau\leq\! c \dm v_\circ\dm_{w(x)}^3,\VSE \hskip 2.5cm\mbox{ for all }t>0\mbox{ and }t-r^2>0.\ea\ee
Considering the last term on the right-hand side of $M(r)$, applying twice H\"older's inequality, \rf{PIII}, we get
\be\label{PV} \ba {ll}\displ r^{-\frac{13}4}\hskip-0.2cm\intll{t-r^2}t\hskip-0.15cm\Big[\intl{|x-y|<r} \hskip-0.3cm|\pi_v(\tau,y)|dy\Big]^\frac54\!d\tau\hskip-0.3cm&\displ\leq 
cr^{-\frac 13}\hskip-0.2cm\intll{t-r^2}t
\Big[\dm \frac{\pi_v(\tau) }{|x-y|}\null_{_{\!\!\frac43}}\!\dm_{\frac32}\Big]^\frac54d\tau\VSE\displ\leq cr^{-\frac13}\hskip-0.2cm\intll{t-r^2}t
\!\dm \frac{v(\tau)}{|x-y|
}\null_{\null_{\frac12}}\!\dm_2^\frac56 \dm\frac{\nabla v(\tau)}{|x-y|}\null_{\null_{\!\frac12}}\!\dm_2^\frac53d\tau\VSE\leq c \dm v_\circ\dm_{w(x)}^\frac52, \hskip 0.5cm\mbox{ for all }t>0\mbox{ and }t-r^2>0.\ea\ee
Hence  \rf{PIV} and \rf{PV} imply, for a suitable constant $C_2$, that
$$M(r)\leq C_2(\mathscr E_0^3+\mathscr E_0^\frac 52).$$ The lemma is proved.\chiu
\section{Proof of Proposition\,\ref{WER}}
The aim of this section is to prove Proposition\,\ref{WER}. We start with following result:
\begin{lemma}\label{CD}{\sl For all suitable weak solutions $(v,\pi_v)$, for $s=0$ and almost everywhere in $s\geq0$,  we get
\be\label{CDI}\lim_{t\to s^+}\intl{\R^3}|v(t,x)-v(s,x)|^2\phi dx=0,\mbox{ for all  }\phi \in L^\infty(\R^3)\,.\ee}\end{lemma}
\Pr It is well known that any weak solution, which satisfies the energy relation in the form i) of Definition\,\ref{WS}, is right-continuous with values in $L^2(\R^3)$, for $s=0$ and almost everywhere in $s\geq0$. Hence, the limit property easily follows.\chiu
\begin{lemma}\label{LWEI}{\sl Under the assumption of Proposition\,\ref{WER},    a suitable weak solution $(v,\pi_v)$   satisfies the following weighted energy inequality:
\be\label{WEIA}\intl{\R^3}\frac{|v(t,y)|^2}{(|x-y|^2+\mu^2)}\null_{\!\frac 12}dy
+c({\mathscr E_0})\intll0t\intl{\R^3}\frac{|\nabla v(\tau,y)|^2}{(|x-y|^2+\mu^2)}\null_{\!\frac 12}dyd\tau\leq \intl{\R^3}\frac{|v_\circ(y)|^2}{(|x-y|^2+\mu^2)}\null_{\!\frac 12},\ee for all $t>0$, $\mu>0$,  and a.e. in $x\in B(x_0,R_0)$.}\end{lemma}
\Pr By virtue of our assumptions, almost everywhere in $x\in B(x_0,R_0)$, we have
\be\label{WEIIA}\intl{\R^3}\frac{|v_\circ(y)|^2}{(|x-y|^2+\mu^2)}\null_{\!\frac12}<c\mathscr E_0^2.\ee
We define $\phi(\tau,y):=(|x-y|^2+\mu^2)^{-\frac12}h_R(y)k(\tau)\in C_0^\infty(\R\times\R^3)$, with $h_R$ and $k$    such that
$$h_R(y):=\left\{\ba{ll}1&\mbox{if }|y|\leq R\\\in (0,1)&\mbox{if }|y|\in [R,2R]\\0&\mbox{for }|y|\geq 2R
\ea\right.\mbox{ and }k(\tau):=\left\{\ba{ll}1&\mbox{if }|\tau|\leq t\\\in (0,1)&\mbox{if }|\tau|\in [t,2t]\\0&\mbox{for }|\tau|\geq 2t
\ea\right.
\,. $$ Substituting this $\phi$ in \rf{SEI}, we get
\be\label{WEIIIA}\ba{l}\displ \intl{\R^3}\!\frac{|v(t,y)|^2h(y)}{(|x-y|^2\!+\!\mu^2)}\null_{\!\frac12}dy
+\!2\!\intll0t\!\!\intl{\R^3}\!\frac{|\nabla v(\tau,y)|^2h(y)}{(|x-y|^2\!+\!\mu^2)}\null_{\!\frac12}dyd\tau\!+\!3\mu^2\!\!\!\intll0t \!\!\intl{\R^3}\!
\frac{|v(\tau,y)|^2h(y)}{(|x-y|^2\!+\!\mu^2)}\null_{\!\frac52}dyd\tau
\VS\hskip1cm\leq
\intl{\R^3}\!\frac{|v_\circ(y)|^2h(y)}{(|x-y|^2\!+\!\mu^2)}\null_{\!\frac12}dy
+\!\intll0t\!\intl{\R^3}\!\frac{|v(\tau,y)|^2h(y)\,v(y)\!\cdot(x-y)}{(|x-y|^2+\mu^2)^\frac 32}\,
dyd\tau\VS\hskip1.5cm+\intll0t\!\intl{\R^3}\!
\frac{\pi_{v}(y)h(y)v(y)\!\cdot\!(x-y)}{(|x-y|^2+\mu^2)^\frac 32}\,dyd\tau+o(R)\VS\hskip 4cm:=\intl{\R^3}\!\frac{|v_\circ(y)|^2h(y)}{(|x-y|^2\!+\!\mu^2)}\null_{\!\frac12}dy+I_1(t,x)+I_2(t,x)+o(R),\ea\ee where
$$\ba{l}\displ o(R):=\intll0t \intl{\R^3}|v|^2\Big[2\nabla h_R\cdot\nabla (|x-y|^2+\mu^2)^{-\frac12}+\frac{\Delta h_R}{(|x-y|^2+\mu^2)}\null_{\!\frac12}+\frac{v\cdot \nabla h_R}{ (|x-y|^2+\mu^2)}\null_{\frac12}\Big]dyd\tau\VS\hskip3cm+\intll0t \intl{\R^3}\frac{\pi_vv\cdot\nabla h_R}{(|x-y|^2+\mu^2)}\null_{\!\frac12}dyd\tau \,.\ea$$ We estimate the terms $I_i,\,i=1,2$. Since $\mu>0$, by virtue of the integrability properties of a suitable weak solution, applying H\"oder's inequality and Lemma\,\ref{WI} we get
$$|I_1(t,x)|\leq 
  \dm \frac{v}{(|x-y|^2+\mu^2)}\null_{\!\frac13}\dm_3 ^3 \leq c
\dm \frac{v}{(|x-y|^2+\mu^2)}\null_{\frac14}\dm_2
\dm \frac{\nabla v}{(|x-y|^2+\mu^2)}\null_{\frac14}\dm_2^2.$$  
For $I_2$ applying the H\"older's inequality and 
Lemma\,\ref{CZW}, we obtain
$$|I_2(t,x)|\leq c\dm \frac{v}{(|x-y|^2+\mu^2)}\null_{\frac13}\dm_3
\dm \frac{\pi_{v}}{(|x-y|^2+\mu^2)}\null_{\frac23}\dm_\frac32\leq c \dm \frac{v}{(|x-y|^2+\mu^2)}\null_{\frac13}\dm_3^3.$$ Hence, as in the previous case, applying  Lemma\,\ref{WI}, we get $$|I_2(t,x)|\leq c
\dm \frac{v}{(|x-y|^2+\mu^2)}\null_{\frac14}\dm_2
\dm \frac{\nabla v}{(|x-y|^2+\mu^2)}\null_{\frac14}\dm_2^2.$$
We increase the right hand side of \rf{WEIIA} by employing the estimates obtained for $I_i,i=1,2$. Hence, by 
applying the Lebesgue dominate convergence theorem, in the limit as $R\to\infty$, for all $t>0$ we deduce the  inequality
\be\label{DIWE}\ba{l}\displ\intl{\R^3}\frac{|v(t,y)|^2}{(|x-y|^2+\mu^2)}\null_{\!\frac12}dy
+2\intll0t\intl{\R^3}\frac{|\nabla v(\tau,y)|^2}{(|x-y|^2+\mu^2)}\null_{\!\frac12}dyd\tau\leq \intl{\R^3}\frac{|v_\circ(y)|^2}{(|x-y|^2+\mu^2)}\null_{\!\frac12}dy\VS\hskip 4cm+c \intll0t\dm \frac{v(\tau)}{(|x-y|^2+\mu^2)}\null_{\frac14}\dm_2
\dm \frac{\nabla v(\tau)}{(|x-y|^2+\mu^2)}\null_{\frac14}\dm_2^2d\tau.\ea\ee
Since, by virtue of Lemma\,\ref{CD}, for all $\mu>0$ $ \frac{v(t,y)}{(|x-y|^2+\mu^2)}\null_{\!\frac12}$ is right-continuous with values in $L^2(\R^3)$ for $s=0$ and almost everywhere in $s\geq0$,  then, $ \mbox{a.e. in
}x\in B(x_0,R_0)$, there exists a
$ [0,\delta(\mu,x))$ in which
\be\label{PB}\intl{\R^3}\frac{|v(t,y)|^2}{(|x-y|^2+\mu^2)}\null_{\!\frac12}dy \leq c \mathscr E_0^2,
\,t\in[0,\delta)\,.\ee Hence, on $[0,\delta)$ estimate \rf{DIWE} becomes 
\be\label{DIWEI}\displ\intl{\R^3}\!\!\frac{|v(t,y)|^2}{(|x-y|^2+\mu^2)}\null_{\!\frac12}dy
+(2-c\mathscr E_0)\!\!\intll0t\!\intl{\R^3}\!\!\frac{|\nabla v(\tau,y)|^2}{(|x-y|^2+\mu^2)}\null_{\!\frac12}dyd\tau\leq
 \intl{\R^3}\!\!\frac{|v_\circ(y)|^2}{(|x-y|^2+\mu^2)}\null_{\!\frac12}dy .\ee
Estimate \rf{DIWEI} implies
 $$ \intl{\R^3}\frac{|v(t,y)|^2}{(|x-y|^2+\mu^2)}\null_{\!\frac 12}dy\leq\intl{\R^3}\frac{|v_\circ(y)|^2}{(|x-y|^2+\mu^2)}\null_{\!\frac 12}dy< c\mathscr E_0^2.$$
Let us prove that this last estimate holds for all $t>0$. Firstly, let us explicitly note that, for all $\mu>0$, the function  $$f(t)=c \intll0t\dm \frac{v(\tau)}{(|x-y|^2+\mu^2)}\null_{\frac14}\dm_2
\dm \frac{\nabla v(\tau)}{(|x-y|^2+\mu^2)}\null_{\frac14}\dm_2^2d\tau$$ is uniformly continuous. Hence there exists $\eta>0$ such that
$$|t_1-t_2|<\eta\Rightarrow |f(t_1)-f(t_2)|< c\mathscr E_0^2-\intl{\R^3}\frac{|v_\circ(y)|^2}{(|x-y|^2+\mu^2)}\null_{\!\frac 12}dy.$$ We claim that estimate \rf{PB} holds in $[\delta,\delta+\eta)$. Assuming the contrary, there exists $\ov t\in [\delta,\delta+\eta)$ such that
\be\label{CONTR}\intl{\R^3}\frac{|v(\ov t,y)|^2}{(|x-y|^2+\mu^2)}\null_{\!\frac 12}dy>c\mathscr E_0^2.\ee On the other hand, the validity of \rf{DIWE} yields 
$$\ba{l}\displ\intl{\R^3}\frac{|v(\ov t,y)|^2}{(|x-y|^2+\mu^2)}\null_{\!\frac12}dy
+2\intll0{\ov t}\intl{\R^3}\frac{|\nabla v(\tau,y)|^2}{(|x-y|^2+\mu^2)}\null_{\!\frac12}dyd\tau\leq \intl{\R^3}\frac{|v_\circ(y)|^2}{(|x-y|^2+\mu^2)}\null_{\!\frac12}dy\VS\hskip 4cm+(f(\ov t)-f(\delta))+f(\delta).\ea$$ Since  via \rf{PB} we get $$f(\delta)\leq c\mathscr E_0\intll0\delta 
\intl{\R^3}\frac{|\nabla v(\tau,y)|^2}{(|x-y|^2+\mu^2)}\null_{\!\frac12}dyd\tau<c\mathscr E_0\intll{0}{\ov t}\intl{\R^3}\frac{|\nabla v(\tau,y)|^2}{(|x-y|^2+\mu^2)}\null_{\!\frac12}dyd\tau,$$ employing the uniform continuity condition, we find 
$$\ba{l}\displ\intl{\R^3}\!\!\frac{|v(\ov t,y)|^2}{(|x-y|^2+\mu^2)}\null_{\!\frac12}dy
+(2-c\mathscr E_0)\!\!\intll0{\ov t}\!\intl{\R^3}\!\!\frac{|\nabla v(\tau,y)|^2}{(|x-y|^2+\mu^2)}\null_{\!\frac12}dyd\tau\VS\hskip3cm\leq
 \intl{\R^3}\!\!\frac{|v_\circ(y)|^2}{(|x-y|^2+\mu^2)}\null_{\!\frac12}dy+f(\ov t)-f(\delta)< c\mathscr E_0^2,\ea$$
which contradicts \rf{CONTR}. Since the arguments are independent of $\delta$, 
we have proved estimate \rf{WEIA}.\chiu
\begin{coro}\label{WEIL}{\sl Under the assumption of Proposition\,\ref{WER}, almost every where in $x\in B(x_0,R_0)$, we get
\be\label{WEILI}\intl{\R^3}\frac{|v(t,y)|^2}{|x-y|}dy+c({\mathscr E_0})\intll0t\intl{\R^3}\frac{|\nabla v(\tau,y)|^2}{|x-y|}dyd\tau\leq c\intl{\R^3}\frac{|v_\circ|^2}{|x-y|}dy,\mbox{ for all }t>0\,.\ee }\end{coro}
\Pr  The thesis is an easy consequence of the following remark:
the families of functions 
$$\big\{\intll0t\intl{\R^3}\frac{|\nabla v(t,y)|^2}{(|x-y|^2+\mu^2)}\null_\frac12dy\big\}\mbox{ and }\big\{\intl{\R^3}\frac{|v(t,y)|^2}{(|x-y|^2+\mu^2)}\null_\frac12dy\big\}$$ are monotone in $\mu>0$. Hence, by virtue of the Beppo Levi's theorem, in the limit as $\mu\to0$, we deduce \rf{WEILI}.\chiu

 \section{Proof of Theorem\,\ref{CT}}
We consider a Lebesgue point $(t,x)$ with $x\in D$ such that \rf{WERI} holds. Let us consider the  parabolic cylinder $Q_{\sqrt t}(\frac76t,x)$. Since our assumptions on $v_\circ$ and $x$ ensure that Proposition\,\ref{WER} holds, by virtue of Lemma\,\ref{RLI} and Theorem\,\ref{CKNT} we get 
$$|v(\tau,y)|\leq 2(c_1)^\frac12t^{-\frac12},\mbox{ in } Q_{\frac {\sqrt t}2}(\mbox{$\frac76t$},x),$$
provided that $(\tau,y)$ is Lebesgue point. Since we are referying to $(t,x)$ which belongs to $ Q_{\frac{\sqrt t}2}(\frac76t,x)$ and recalling the expression of $c_1=c_0(\mathscr E_0^3+\mathscr E_0^\frac52)^\frac23$, we have proved the theorem.\chiu
\vskip0.1cm
 {\bf Acknowledgments} - The result of this paper was part of a broader paper, on the asymptotic behavior of a suitable weak solution (see \cite{CMSD}), and mainly represents its natural premise. In this connection the authors thank Professor P. Deuring: after a private conversation of the second author, where P. Deuring  reaffirmed the absence of a result of continuous dependence in Corollary on page 820 in \cite{CKN}, and also the interest of this question, the authors considered worthwhile to extract this note from paper \cite{CMSD}. \par This research was partly supported by GNFM-INdAM, and by MIUR via the PRIN 2012 {\it ``Nonlinear Hyperbolic Partial Differential Equations, Dispersive and Transport Equations: Theoretical and Applicative Aspects''}.


\begin{thebibliography}{20}
\bibitem{CKN}L. Caffarelli, R. Kohn and L. Nirenberg, {\it Partial regularity of suitable weak solutions of the Navier-Stokes equations}, 
Comm. Pure Appl. Math., {\bf  35} (1982), 771-831. 
\bibitem{CMSD} F. Crispo and P. Maremonti, {\it On the spatial asymptotic decay of a suitable weak solution to the Navier-Stokes Cauchy problem}, submitted. 
 \bibitem{Farwig} R. Farwig, {\it Partial regularity and weighted energy estimates of global weak solutions of the Navier-Stokes system},
  Progress in partial differential equations: the Metz surveys, 4, 205--215, Pitman Res. Notes Math. Ser., 345, Longman, Harlow, 1996. 
  
 \bibitem{LS}O.A. Ladyzhenskaya and  G.A. Seregin, {On partial regularity of suitable weaks olutions to the three-dimensional Navier-Stokes equations}, J. Math. Fluid Mech., {\bf 1}  (1999), 356-387.
 \bibitem{L} F. Lin, {\it A new proof of the Caffarelli-Kohn-Nirenberg theorem}, Comm. Pure Appl. Math., {\bf 51} (1998), 241--257.
\bibitem{MRP}P. Maremonti, {\it Partial regularity of a generalized solution to the Navier-Stokes equations in exterior domain}, Comm. Math. Phys., {\bf 110}  (1987), 75-87.
\bibitem{MCMP} P. Maremonti, {\it On the asymptotic behavior of the $L^2$-norm of suitable weak solutions to the Navier-Stokes equations in three-dimensional exterior domains}, Comm. Math. Phys., {\bf 198} (1988), 385--400.
\bibitem{MSAP}{P. Maremonti and V.A. Solonnikov}, {\it An estimate for the solutions of Stokes equations in exterior domains}, Zap. Nauch. Sem. LOMI, {\bf 180} (1990), 105--120, trasl. in J. Math. Sci., {\bf 68} (1994), 229--239.
\bibitem{MSMS}{P. Maremonti and V.A. Solonnikov}, {\it On nonstationary Stokes problem in exterior domains}, Ann. Sc. Norm. Sup. Pisa, {\bf 24} (1997), 395--449. 
\bibitem{J} J.A. Mauro, {\it Some analytic questions in mathematical physic problems}, Pliska Stud. Math. Bulgar., (2014). \bibitem{scheffer1} V. Scheffer, {\it Hausdorff measure and the Navier-Stokes equations}, Comm. Math. Phys., {\bf 55} (1977),  97--112.
\bibitem{S}G.A. Seregin, {\it Local regularity for suitable weak solutions of the Navier-Stokes equations}, Russian Math. Surveys, {\bf 62}   (2007), 595--614.
\bibitem{ST}{E.A. Stein}, {\it Note on singular integrals}, Proc. Amer. Math. Soc., {\bf 8} (1957), 250--254.
\bibitem{V}A. Vasseur, {\it A new proof of partial regularity of solutions to Navier-Stokes equations}, Nonlin. Diff. Eq. Appl., {\bf 14} (2007), 753--785.
\end{thebibliography}
\end{document}